\documentclass[dvips]{acta}
\usepackage{supertabular,lscape,epsfig}
\usepackage{amssymb}
\usepackage{amsmath}
\DeclareSymbolFont{ppa}{OT1}{ppl}{m}{it}
\DeclareMathSymbol{\vv}{\mathalpha}{ppa}{'166}

\newfont{\hb}{rphvb at 10pt}%bezszeryfowe pó³grube
\newfont{\hbo}{rphvbo at 10pt}%bezszeryfowe pó³grube kursywa
\newfont{\bitt}{rptmbi at 12pt}%pó³gruba kursywa (tytu³ artyku³u)
\newfont{\bits}{rptmbi at 11pt}%pó³gruba kursywa (tytu³y rozdzia³ów)

\SetPages{179}{196}

\SetVol{60}{2010}

\begin{document}

%Zwarte naglowki, jeden wiersz
\newcommand{\TabCapp}[2]{\begin{center}\parbox[t]{#1}{\centerline{
  \small {\spaceskip 2pt plus 1pt minus 1pt T a b l e}
  \refstepcounter{table}\thetable}
  \vskip2mm
  \centerline{\footnotesize #2}}
  \vskip3mm
\end{center}}

%Zwarte naglowki, dwa wiersze
\newcommand{\TTabCap}[3]{\begin{center}\parbox[t]{#1}{\centerline{
  \small {\spaceskip 2pt plus 1pt minus 1pt T a b l e}
  \refstepcounter{table}\thetable}
  \vskip2mm
  \centerline{\footnotesize #2}
  \centerline{\footnotesize #3}}
  \vskip1mm
\end{center}}

%Zwarte naglowki, jeden wiersz
\newcommand{\MakeTableSepp}[4]{\begin{table}[p]\TabCapp{#2}{#3}
  \begin{center} \TableFont \begin{tabular}{#1} #4 
  \end{tabular}\end{center}\end{table}}

%Zwarte naglowki, jeden wiersz
\newcommand{\MakeTableee}[4]{\begin{table}[htb]\TabCapp{#2}{#3}
  \begin{center} \TableFont \begin{tabular}{#1} #4
  \end{tabular}\end{center}\end{table}}

%Zwarte naglowki, dwa wiersze
\newcommand{\MakeTablee}[5]{\begin{table}[htb]\TTabCap{#2}{#3}{#4}
  \begin{center} \TableFont \begin{tabular}{#1} #5 
  \end{tabular}\end{center}\end{table}}

%FWHM, PSF - proste, MgII, H$\alpha$
%rms, rhs, sd - kursywa
%{\sc DAOPhot}
%{\sf files}
%Galactic wszystko (bulge, center, plane...)
%Cepheids
%type~ Cepheids, Population~II Cepheids
%light curve
\newfont{\bb}{ptmbi8t at 12pt}
\newfont{\bbb}{cmbxti10}
\newfont{\bbbb}{cmbxti10 at 9pt}
\newcommand{\uprule}{\rule{0pt}{2.5ex}}
\newcommand{\douprule}{\rule[-2ex]{0pt}{4.5ex}}
\newcommand{\dorule}{\rule[-2ex]{0pt}{2ex}}
\def\thefootnote{\fnsymbol{footnote}}
\begin{Titlepage}
\Title{The Optical Gravitational Lensing Experiment.\\
The OGLE-III Catalog of Variable Stars.\\
X.~Enigmatic Class of Double Periodic Variables\\ 
in the Large Magellanic Cloud
\footnote{Based on observations obtained with the 1.3-m 
Warsaw telescope at the Las Campanas Observatory of the Carnegie
Institution of Washington.}}
\Author{R.~~P~o~l~e~s~k~i$^1$,~~
I.~~S~o~s~z~y~ñ~s~k~i$^1$,~~ 
A.~~U~d~a~l~s~k~i$^1$,~~ 
M.\,K.~~S~z~y~m~a~ñ~s~k~i$^1$,\\ 
M.~~K~u~b~i~a~k$^1$,~~ 
G.~~P~i~e~t~r~z~y~ñ~s~k~i$^{1,2}$,~~ 
£.~~W~y~r~z~y~k~o~w~s~k~i$^3$~~ and~~ 
K.~~U~l~a~c~z~y~k$^1$}
{$^1$ Warsaw University Observatory, Al. Ujazdowskie 4, 00-478 Warszawa,
Poland\\ 
e-mail:
(rpoleski,soszynsk,udalski,msz,mk,pietrzyn,kulaczyk)@astrouw.edu.pl\\
$^2$ Universidad de Concepción, Departamento de Fisica, Casilla 160-C,
Concepción, Chile\\ 
$^3$ Institute of Astronomy, University of Cambridge, Madingley Road, Cambridge CB3 0HA, UK\\
e-mail: wyrzykow@ast.cam.ac.uk}
\Received{September 15, 2010}
\end{Titlepage}

\Abstract{
The tenth part of the OGLE-III Catalog of Variable Stars contains 125
Double Periodic Variables (DPVs) from the Large Magellanic Cloud. DPVs are
semi-detached binaries which show additional variability with a period
around 33 times longer than the orbital period. The cause of this long
cycle is not known and previous studies suggest it involves circumbinary
matter.

We discuss the properties of the whole sample of the LMC DPVs and put more
attention to particularly interesting objects which may be crucial for
verifying hypothesis explaining long cycle variability. Secondary eclipses
of one of the objects disappear during some orbital cycles and primary
eclipses are deeper during long cycle minimum.}{Catalogs -- binaries: close
-- Magellanic Clouds}

\Section{Introduction}
Double Periodic Variables (DPVs) were first recognized as a separate class
of variable stars by Mennickent \etal (2003). Their inspection of the
photometry collected during the second phase of the Optical Gravitational
Lensing Experiment (OGLE) revealed a group of 27 Large Magellanic Cloud
(LMC) and 3 Small Magellanic Cloud (SMC) blue stars which simultaneously
show two periods which ratio is close to 35.2. The light curve of the
shorter period ($P_1$) in some instances was characteristic for eclipsing
binaries. The longer periods ($P_2$) were in the interval between 140 and
960 days and the shape of the light curves was sinusoidal in most cases. It
was speculated that the latter periods were caused by the precession of an
elliptical disk around the blue component of the semi-detached
binary. Mennickent \etal (2005) revealed three new DPVs in the LMC and
analyzed the spectroscopic observations done for some of the systems. The
multi-epoch spectroscopy of two DPVs with small amplitudes and sinusoidal
light curves of the short period shows that their brightness changes are
caused by the ellipsoidal variation of one of the components of the binary
system. This finding was crucial for assigning short periods or twice
larger values of all DPVs to orbital periods of binary
systems. Mennickent \etal (2005) also had an impression that the longer
cycle variability disappears during the main eclipse, this finding was
negated later on. As a consequence, this variability should be caused by a
phenomenon taking place in or near the surface of brighter
component. However, any of the known types of stellar variability is in the
agreement with observed features. Another important finding made by
Mennickent \etal (2005) was a period shortening and an amplitude increase
in the long period of one LMC DPV. Buchler \etal (2009) re-investigated
MACHO project photometry for 30 DPVs selected by Mennickent
\etal (2003) and revealed that after prewhitening with the two most prominent
frequencies also a sum of these frequencies is significant\footnote{Note
that Buchler \etal (2009) used different frequencies designations in the
text and in their Table~1.}. No further discussion of this combination
frequency is given in the literature.

Mennickent \etal (2008) observed spectroscopically an eclipsing DPV in the
LMC. Their investigation resulted in a model which contains a binary system
with the Roche lobe-filling secondary, hotter primary star with the
circumprimary disk, mass transfer and mass outflow feeding the circumbinary
disk. Contrary to the previous studies, it was inferred that the source of
the longer cycle is in the circumbinary matter as no interference of short
and long period was found. Desmet \etal (2010) used space mission CoRoT
photometry and high-resolution ground-based spectroscopy to investigate
Galactic DPV star AU Mon. CoRoT photometry was obtained during minimum of
the longer cycle. Their conclusions regarding the model of the binary are
consistent with previous studies of other DPVs. High quality space-based
photometry allowed finding two additional periods in residual light
curve. Both these periods are approximately a hundred times shorter than
the orbital period of the binary. Djura\v{s}eviæ \etal (2010) deepened the
analysis of CoRoT photometry and added a hot spot and two bright spots to
the models of the optically and geometrically thick disk. They explained
the period-to-period changes in the CoRoT light curve by a variable
contribution of the disk brightness. This changes cannot explain the long
term cycle and Djura\v{s}eviæ \etal (2010) concluded that the circumbinary
matter has to be responsible for the long cycle variability. Another
Galactic DPV -- V393 Sco -- was spectroscopically observed by Mennickent
\etal (2010). They favor equatorial mass loss through the Lagrangian point
$L_3$ and argue against the polar jets as a source of the longer cycle. The
velocity of the mass lost through $L_3$ was estimated to be about 300~km/s.

Finding variability with a period of a few hundred days and amplitude of
about 0.1~mag superimposed on the variability caused by a binary system
with much shorter period is possible when the long-term uniformly obtained
photometry is available. That is why almost the whole photometry of DPVs
analyzed in the papers mentioned above came from the OGLE or the MACHO
project -- two microlensing surveys monitoring the LMC and the SMC. In this
paper we use OGLE-III photometry to search for DPVs in the LMC. We increase
the number of published stars of this type almost by a factor of four. Our
search yields not only the higher number of DPVs in the LMC, which allows
better statistical description of this group of stars, but also we present
several objects with unexpected features. They may be crucial in
understanding the cause of long term cycle, even though, we do not provide
any self-consistent model explaining all observed features of the long term
cycle.

In the next sections we discuss observations and the procedure of selecting
DPVs. Section~4 presents the catalog itself and is followed by a discussion
of properties of the whole sample as well as selected objects. We end with
a summary and future prospects. To avoid confusion, we denote phases of the
shorter (orbital) and longer cycle as $\phi_1$ and $\phi_2$,
respectively. In the case of the orbital cycle $\phi_1=0$ corresponds to
the minimum light (primary eclipse in eclipsing systems), while for the
longer cycle $\phi_2=0$ corresponds to the maximum light, which is more
distinctive in most cases. Similarly, frequencies $f_1$ and $f_2$ are equal
$1/P_1$ and $1/P_2$, respectively.
\vspace*{-9pt}
\Section{Observations}
\vspace*{-5pt}
The OGLE-III project observed the LMC between June 2001 and May 2009 and
covered around 40 square degrees. The observations were conducted with the
1.3~m Warsaw Telescope situated at Las Campanas Observatory, which is
operated by Carnegie Institution of Washington. The telescope was equipped
with the eight chip mosaic camera with the total dimension
8k$\times$8k~pixels. The pixel size of 15~$\mu$m gives the pixel scale of
0\zdot\arcs26 and the total field of view around
$35\arcm\times35\arcm$. The detailed description of the instrumentation can
be found in Udalski (2003). The photometry was performed using Difference
Image Analysis technique (Alard and Lupton 1998, Alard 2000, Wo¼niak 2000)
which works very well in crowded fields. Udalski \etal (2008a) give full
description of the photometric and astrometric data reduction process. The
DIA photometric uncertainties are known to be underestimated and thus were
corrected using the method described by Wyrzykowski \etal (2009).

Typically, there are 500 photometric measurements per star obtained during
OGLE-III observations. Around 90\% of them were secured using {\it I}
filter. The rest was taken in the {\it V}-band. For stars in the central
parts of the LMC we connected OGLE-III and OGLE-II photometry (Szymañski
2005) obtained from 1997 to 2000. To keep the data in the same photometric
system we added to the OGLE-II magnitudes the difference between median
OGLE-III and OGLE-II magnitudes. In some cases with a different sampling of
the short or long cycles, additional correction found manually was
added. Similar procedure was carried out for objects present in two or
three OGLE-III fields in the overlapping parts of adjacent fields. Our
final photometry contains up to 1500 {\it I}- and 270 {\it V}-band
measurements. One may find {\it B}-band mean magnitudes for stars in
central parts of the LMC in Udalski \etal (2000).
\vspace*{-9pt}
\Section{Selection of DPVs}
\vspace*{-5pt}
We have searched for characteristic period ratio in the results of the
massive period search done for OGLE-III photometry of all stars observed in
the LMC by the OGLE-III survey (Soszyñski \etal 2008). More intense search
was performed for stars in the region of the color--magnitude diagram where
DPVs are expected ($V-I<0.6$~mag and $V<19$~mag; Mennickent \etal
2005). For these blue objects we used prewhitening with a variety of a
different number of Fourier harmonics. In all cases the stars with the
period ratios in a broad region around expected value of 35, even two times
larger or smaller, were visually examined. Altogether 113 DPVs were
selected in this way.

We suspected that our searches still may not reveal all the DPVs because
automatic search for eclipsing binary periods may give spurious results
(Derekas \etal 2007). Fourier fitting procedure with fallacious number of
harmonics may also cause some artifacts in prewhitened light curve
hampering searches for the additional longer period. Because of that we
decided to visually inspect light curves of automatically selected
eclipsing binaries candidates with periods longer than one
day.\footnote{The catalog of Magellanic Clouds eclipsing binaries with
periods longer than one day is under construction by Graczyk \etal (in
preparation).}  26\,000 blue candidates and 7700 candidates without color
information in OGLE-III data were inspected and stars showing light curves
typical for DPVs phased with the orbital period were selected and studied
in more detail. Additional eight stars were added to the list of
DPVs. Another three DPVs were found by authors during other variability
searches.

Our list of DPVs in the LMC was cross-matched with the stars found by
Mennickent \etal (2003) and Mennickent \etal (2005). One of their stars was
not detected by our procedure. Its designation is OGLE-LMC-DPV-114 (MACHO
ID: 76.9844.110). This star does not have the {\it V}-band magnitude in the
OGLE-III photometric maps (Udalski \etal 2008b) and it was not among the
eclipsing binary candidates inspected visually. We decided to add it to our
catalog. We note that our selection process was less efficient for stars
with orbital periods shorter than one day as many artifacts appear in this
period range. Also stars with exceptionally large $P_2$ values or ones with
$P_2\approx1$~yr could have been overlooked.

\begin{figure}[p]
\centerline{\includegraphics[height=8cm, bb=20 20 750 560]{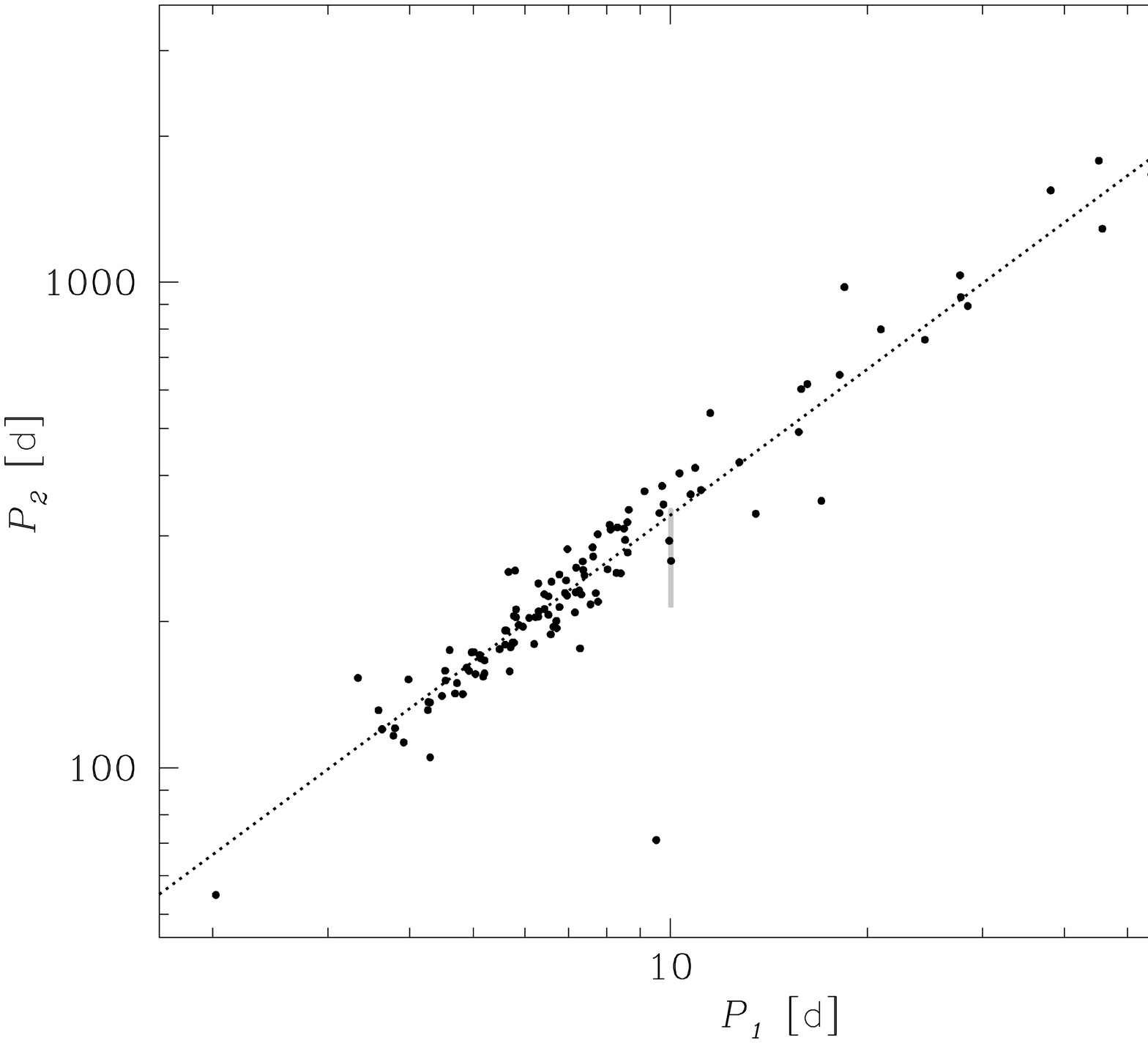}} 
\FigCap{Relation between the orbital and long period. Uncertainties 
of periods are smaller than size of points. Dotted line represents the
relation $P_2=33.13P_1$. Gray vertical line shows the periods of
OGLE-LMC-DPV-065 during the last 18 years. See Section~5.7 for discussion.}
\vskip7pt
\centerline{\includegraphics[height=13cm,angle=270]{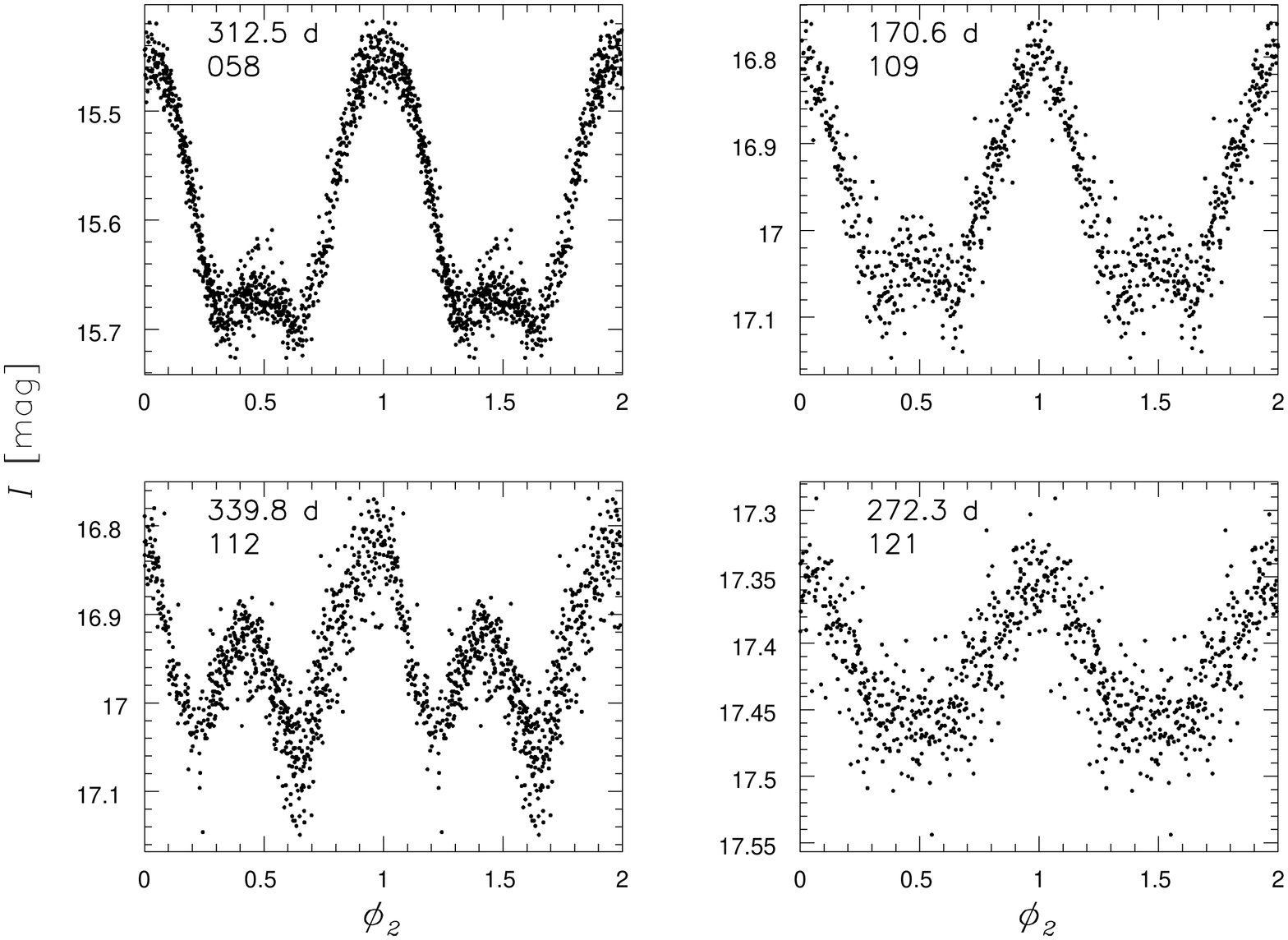}} 
\FigCap{Exemplary non-sinusoidal light curves of long cycles. Values of 
$P_2$ and catalog numbers in the catalog are given in each panel.}
\end{figure}
Fig.~1 shows the $P_1$--$P_2$ relation for all detected DPVs. The median
value of $P_2/P_1$ is 33.13 and the dotted line shows this relation. The
same diagram made using preliminary periods showed twelve stars with $P_2$
values twice smaller than the rest with similar $P_1$. For these objects we
multiplied $P_2$ by a factor of two. Fig.~2 presents four exemplary light
curves showing non-sinusoidal signal or even two maxima in one cycle. They
ensure us that these twelve stars do not differ significantly from other
DPVs if we take $P_2$ twice longer than photometric period. The largest
$P_2$ values are comparable to the time-span of OGLE-III observations so it
is not sure if we observe periodic phenomenon. The values of $P_1$ lie in
the range between 2 and 106 days.

All the periods were verified using software {\sc Tatry} utilizing mhAOV
method described by Schwarzenberg-Czerny (1996). All the light curves were
prewhitened with case-by-case selected number of Fourier harmonics. The
second periods were also checked using mhAOV method. We note that in first
periodograms we see $P_2$ or $P_2/2$ as a dominant twice more often than
$P_1$ or $P_1/2$.

\Section{Catalog}
The OGLE-III catalog of DPVs in the LMC contains 125 objects and,
identically as other parts of the OGLE-III Catalog of Variable Stars, is
available through the OGLE Internet archive both as a user-friendly WWW
interface and from the anonymous FTP site:
\begin{center}
{\it http://ogle.astrouw.edu.pl/\\
ftp://ftp.astrouw.edu.pl/ogle/ogle3/OIII-CVS/lmc/dp$\vv$/}
\end{center}
The FTP site contains file {\sf ident.dat} which successive columns
contain: catalog designation in the form OGLE-LMC-DPV-XXX where XXX is
consecutive number (the objects were sorted in order of increasing right
ascension), two columns giving OGLE-III designation taken from Udalski
\etal (2008b), equinox J2000.0 right ascension and declination,
cross-identification with the OGLE-II and MACHO photometric databases and
cross-identification with the extragalactic part of the General Catalog of
Variable Stars (Artyukhina \etal 1995). File {\sf DPV.dat} gives for each
star: {\it I} and {\it V} magnitudes corresponding to the mean brightnesses
of long cycle and maximum brightnesses of orbital cycle, properties of
orbital variability (period and its uncertainty, epoch of the minimum
light, peak-to-peak amplitude in the {\it I}-band) and properties of the
longer cycle (the same as for the orbital one but with the epoch of the
maximum instead of minimum light). Additional remarks for individual
objects are given in the file {\sf remarks.dat}. The subdirectory {\sf
phot/} contains multi-epoch {\it V}- and {\it I}-band OGLE
photometry. Finding charts are given in the subdirectory {\sf
fcharts/}. The $60\arcs\times60\arcs$ charts are orientated with N up and E
to the left.

\begin{figure}[htb]
%\centerline{\includegraphics[height=12.7cm,angle=270]{fig3.ps}}
\centerline{\includegraphics[height=12.7cm,]{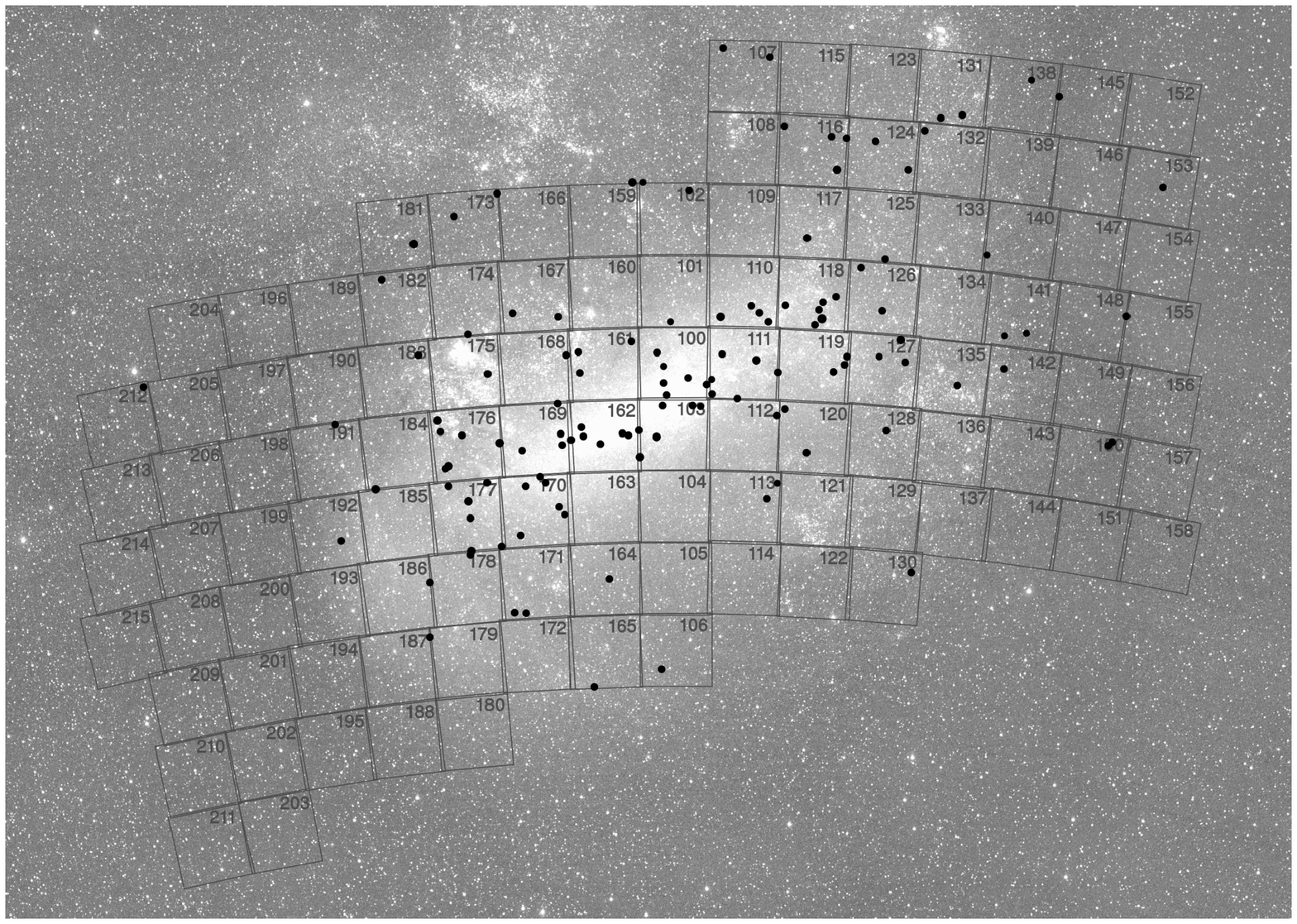}}
\bigskip
\FigCap{Spatial distribution of DPVs in the LMC. Background squares 
show OGLE-III fields. The image of the LMC comes from Pojmañski (1997).}
\end{figure}
Fig.~3 presents the spatial distribution of presented DPVs. Most of them
are found in the LMC bar, thus we do not suspect many of these objects to
lie outside the OGLE-III fields of the LMC.
\vspace*{-9pt}
\Section{Discussion}
\vspace*{-5pt}
In this section we firstly discuss the properties of DPVs as a group and
then describe in details the most interesting individual objects.
\vspace*{-7pt}
\subsection{Color--Magnitude Diagram}
\vspace*{-5pt}
\begin{figure}[htb]
%\centerline{\includegraphics[width=9.5cm, bb=20 150 570 690]{fig4.ps}} 
\centerline{\includegraphics[width=9.3cm, bb=20 20 565 565]{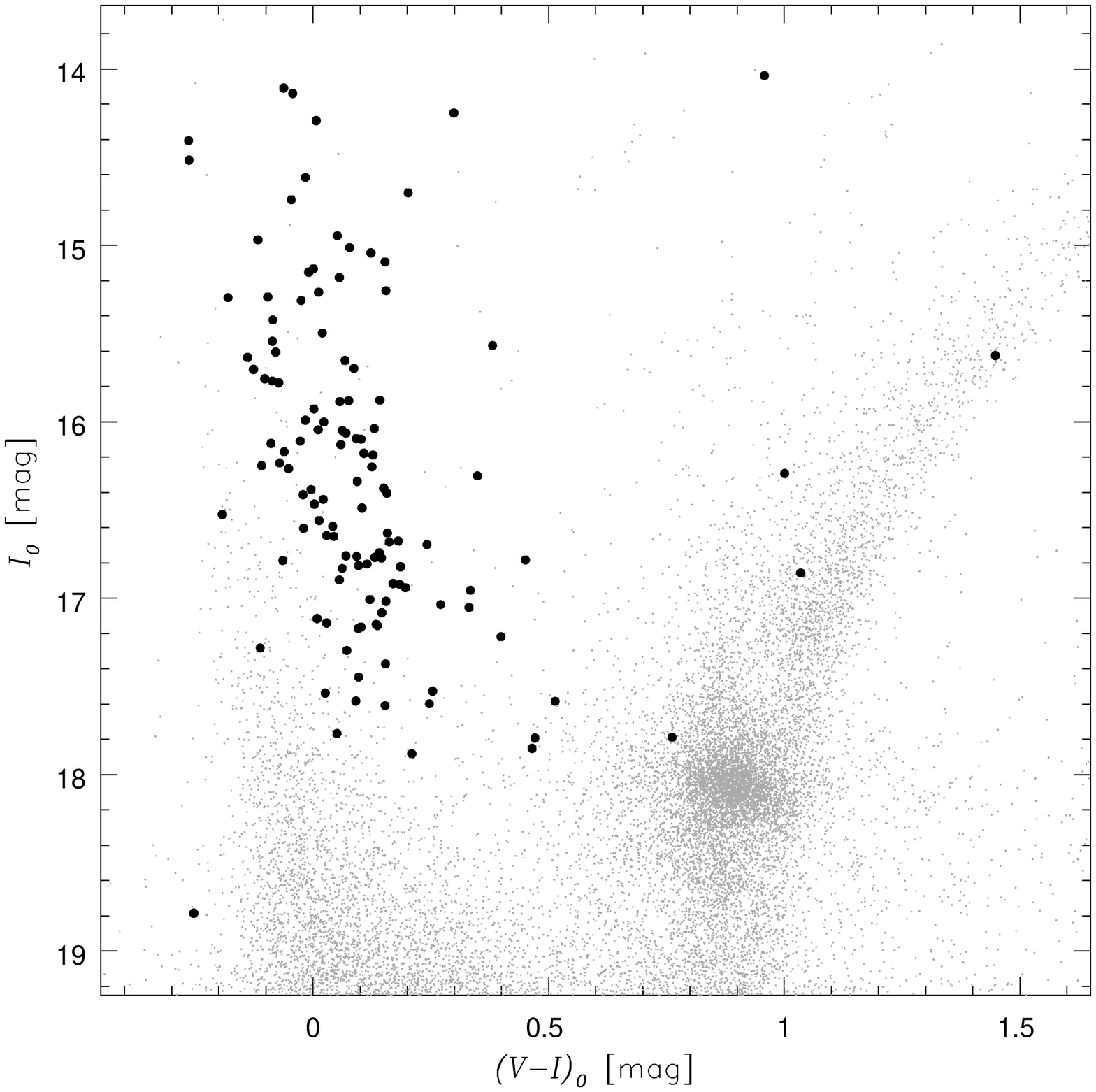}} 
\FigCap{Color--magnitude diagram for DPVs. Gray background 
shows a sample of stars from OGLE-III subfield LMC162.4 for comparison.}
\end{figure}
The color--magnitude diagram (CMD) is shown in Fig.~4. For calculation of
extinction-corrected $\left(V-I\right)_0$ colors and $I_0$ brightnesses we
used Pejcha and Stanek (2009) map. Five of the reddest stars with
$\left(V-I\right)_0>0.75$~mag show light curves at least slightly different
from other DPVs and thus were marked as uncertain. One of them
(OGLE-LMC-DPV-074) shows an interaction of the shorter and longer cycles
what convinced us that both periodicities originate in the same object. It
is characterized below. The short cycle light curve of the faintest DPV is
not a typical light curve of binary system. This object is also marked as
uncertain. All the other DPVs are in the region of CMD restricted by
relations $I_0<17.9$~mag and $\left(V-I\right)_0<0.59$~mag. They are
brighter than the main sequence stars with the same color what is
consistent with their binary nature. Some DPVs may be overexposed in the
OGLE-III images and thus not included in our catalog.

\subsection{Amplitude of the Long Cycle During Minimum of the 
Orbital Cycle}
\begin{figure}[htb]
%\centerline{\includegraphics[width=10cm, bb=20 30 570 745]{fig5.ps}} \
\centerline{\includegraphics[width=9.7cm, bb=20 20 565 745]{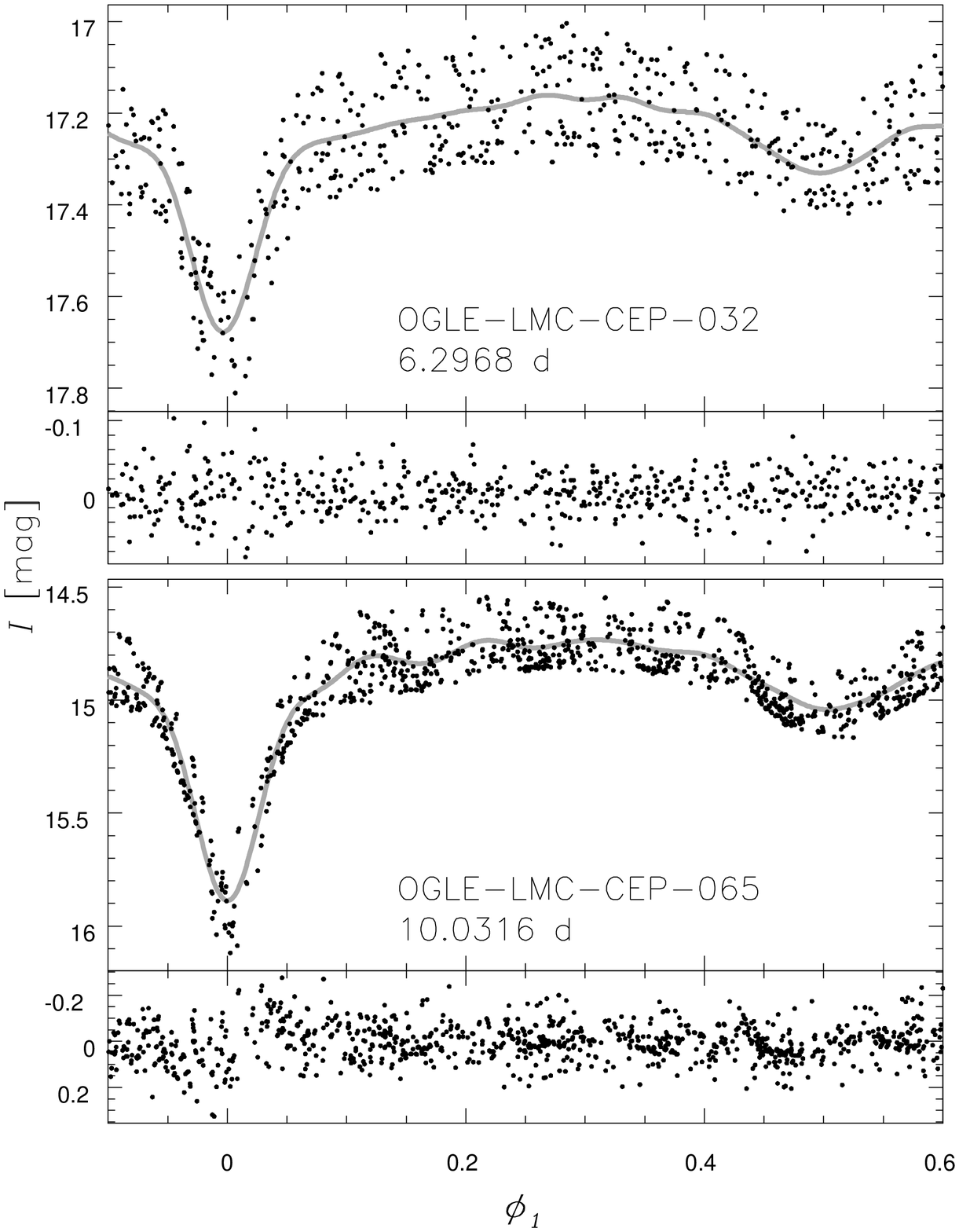}} 
\FigCap{Orbital light curves of two DPVs presented by Mennickent 
\etal (2005). Only phases from $-0.1$ to 0.6 are shown. Designations 
of stars and their orbital periods are given in each panel. Gray lines
represent Fourier fits and residuals are given below each light curve. No
definite dependence of long cycle amplitude on phase of the orbital cycle
is seen.}
\end{figure}

We have visually inspected all the unfiltered light curves phased with the
long and short periods to search for connections between both cycles. We
did not find any such connection except for a few examples mentioned
below. Mennickent \etal (2005) suggested that the long cycle variability
disappears during the primary minimum. We do not confirm this conclusion in
general case. In Fig.~5 we show two light curves with $\phi_1$ limited to
$(-0.1,0.6)$ interval. The stars are the same as shown by Mennickent \etal
(2005) in their Fig.~12. Similarly to Mennickent \etal (2008) we think that
the suggestion of Mennickent \etal (2005) was an effect of the high
inclination of the eclipse branches. Also small number of measurements in
short $\phi_1$ interval might have affected the visual impression.

\subsection{Combination Frequencies}
\MakeTableee{l@{\hspace{3pt}}r@{\hspace{7pt}}l@{\hspace{3pt}}c@{\hspace{3pt}}l@{\hspace{3pt}}rl}{12.5cm}{Combination frequencies}
{\cline{1-3}\cline{5-7}
\noalign{\vskip3pt}
\multicolumn{1}{c}{ID} & \multicolumn{1}{c}{$P_3$ [day]} & \multicolumn{1}{c}{$f_3$} &
$\phantom{xxx}$&
\multicolumn{1}{c}{ID} & \multicolumn{1}{c}{$P_3$ [day]} & \multicolumn{1}{c}{$f_3$} \\ 
\noalign{\vskip3pt}
\cline{1-3}\cline{5-7}
\noalign{\vskip3pt} 
OGLE-LMC-DPV-002 & 4.1640  & $f_1+f_2$    && OGLE-LMC-DPV-077 & 6.5933  & $f_1+f_2$ \\	      
OGLE-LMC-DPV-015 & 8.1342  & $f_1+f_2$    && OGLE-LMC-DPV-078 & 7.0423  & $f_1+f_2$ \\	      
OGLE-LMC-DPV-019 & 7.1568  & $f_1+f_2$    && OGLE-LMC-DPV-080 & 7.0893  & $f_1+f_2$ \\	      
OGLE-LMC-DPV-020 & 8.2769  & $f_1+f_2$    && OGLE-LMC-DPV-083 & 6.4859  & $f_1+f_2$ \\	      
OGLE-LMC-DPV-021 & 7.1819  & $f_1+f_2$    && OGLE-LMC-DPV-084 & 15.4522 & $f_1+f_2$ \\	      
OGLE-LMC-DPV-027 & 6.0406  & $f_1+f_2$    && OGLE-LMC-DPV-088 & 4.4439  & $2(f_1+f_2)$ \\
OGLE-LMC-DPV-029 & 5.0362  & $f_1+f_2$    && OGLE-LMC-DPV-089 & 6.9880  & $f_1+f_2$ \\	      
OGLE-LMC-DPV-034 & 25.3273 & $f_1-f_2$    && OGLE-LMC-DPV-092 & 18.1226 & $f_1+f_2$ \\	      
OGLE-LMC-DPV-035 & 7.4471  & $f_1+f_2$    && OGLE-LMC-DPV-097 & 7.5582  & $f_1+f_2$ \\	      
OGLE-LMC-DPV-038 & 9.5017  & $f_1+f_2$    && OGLE-LMC-DPV-098 & 5.4605  & $f_1+f_2$ \\	      
OGLE-LMC-DPV-047 & 5.6602  & $f_1+f_2$    && OGLE-LMC-DPV-101 & 3.9654  & $2(f_1+f_2)$ \\
OGLE-LMC-DPV-048 & 10.3202 & $f_1-f_2$    && OGLE-LMC-DPV-102 & 5.6999  & $f_1+f_2$ \\	      
OGLE-LMC-DPV-050 & 5.5382  & $f_1+f_2$    && OGLE-LMC-DPV-105 & 7.3005  & $f_1+f_2$ \\	      
OGLE-LMC-DPV-063 & 23.7287 & $2(f_1+f_2)$ && OGLE-LMC-DPV-115 & 7.4150  & $f_1+f_2$ \\	      
OGLE-LMC-DPV-064 & 4.6620  & $f_1+f_2$    && OGLE-LMC-DPV-116 & 5.5946  & $f_1+f_2$ \\	      
OGLE-LMC-DPV-066 & 9.3615  & $f_1+f_2$    && OGLE-LMC-DPV-118 & 5.4874  & $f_1+f_2$ \\	      
OGLE-LMC-DPV-067 & 5.5608  & $f_1+f_2$    && OGLE-LMC-DPV-119 & 4.8723  & $f_1+f_2$ \\	      
OGLE-LMC-DPV-069 & 6.4867  & $f_1+f_2$    && OGLE-LMC-DPV-123 & 6.1941  & $2(f_1+f_2)$ \\
\noalign{\vskip3pt}
\cline{1-3}\cline{5-7}}

All the light curves were prewhitened with both $P_1$ and $P_2$ periods and
periodograms were calculated for residual data. We checked if significant
frequencies found during that search are linear combinations of $f_1$ and
$f_2$. Table~1 presents for each star for which we found combination
frequency the period corresponding to it (column 2) and its equivalence
(column~3). Altogether 36 star are listed which is 29\% of the
sample. Most common are $f_1+f_2$ combinations which are found in 30 cases.

\begin{figure}[htb]
\centerline{\includegraphics[width=9cm]{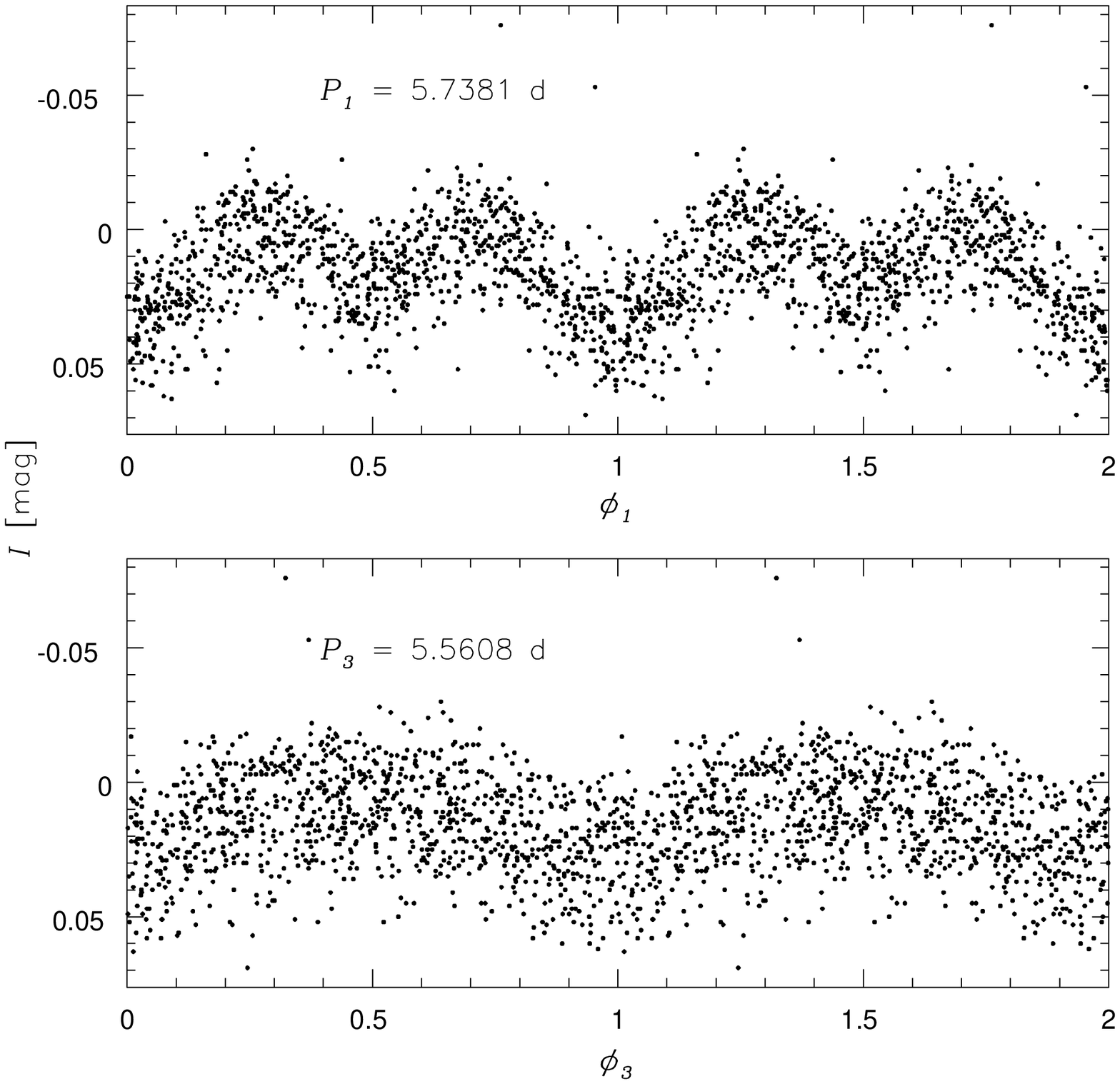}} 
\FigCap{Light curve of OGLE-LMC-DPV-067 prewhitened with 
$P_2=179.96$~days and phased with $P_1$ ({\it upper panel}) and $P_3$
({\it lower panel}). $P_3$ is a combination periodicity.}
\end{figure}
Out of ten LMC DPVs for which Buchler \etal (2009) found combination
frequencies we confirm five cases. They prewhitened data using half of
$P_1$ and $P_2$, we used $P_1$ and $P_2$. The results are very similar --
around 30\% of DPVs show these additional frequencies. Combination
frequencies may emerge after prewhitening with two frequencies if periods
used are inaccurately determined or wrong number of harmonics are used. For
a randomly selected DPV with a combination frequency we show the light
curve after subtraction of long cycle (Fig.~6) phased with both $P_1$ and
$P_3$. The signal with a period $P_3$ is seen before prewhitening with
$P_1$ what is a strong suggestion that $P_3$ is not an artifact.

Except for frequencies reported above we also found other which were not
combinations of $f_1$ and $f_2$. We report them in the remarks to the
catalog.

\subsection{OGLE-LMC-DPV-074 -- Smaller Amplitude of Long Cycle
During\\ Primary Eclipse}
The light curve of this object is quite different from the remaining DPVs
described here. The upper panel of Fig.~7 presents the light curve of
OGLE-LMC-DPV-074 folded with the orbital period. The lower panel shows data
prewhitened with the orbital cycle, and still phased with the orbital
period. One can easily notice that the differences between the fitted
\begin{figure}[htb]
\centerline{\includegraphics[height=7cm, bb=20 20 750 575]{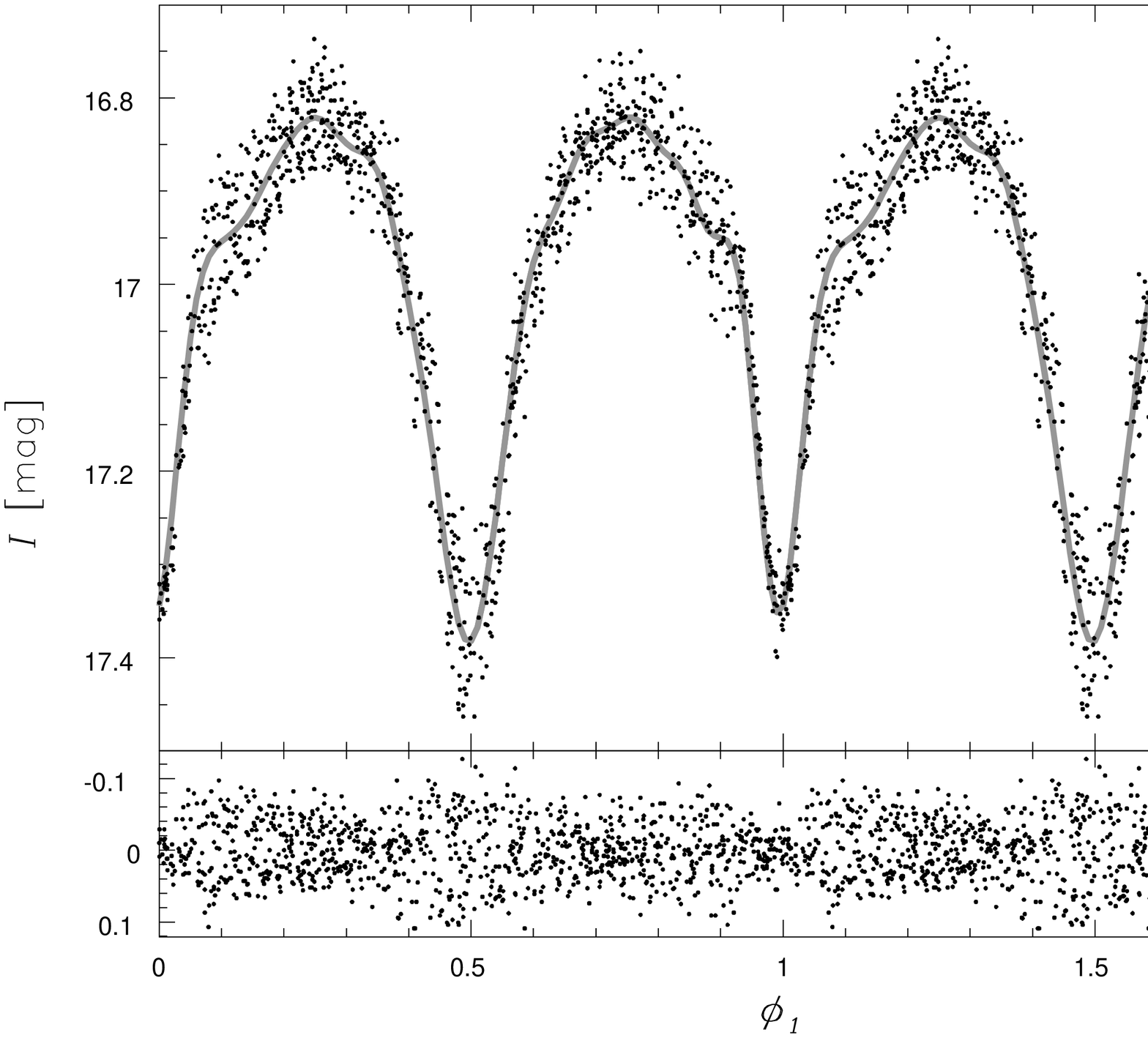}} 
\FigCap{Light curve of OGLE-LMC-DPV-074 ($P_1=38.1592$~days) with the 
fitted Fourier model and phased with the orbital period. {\it Lower panel}
shows residuals. Long cycle activity weakens close to phase~0.}
\end{figure}
Fourier series and the actual measurements are much smaller close to
$\phi_1=0$ than close to $\phi_1=0.5$. It seems that the amplitude of the
longer cycle gets smaller close to $\phi_1=0$. In fact we are not sure
which of the eclipses is primary and which one is secondary, because the
depth of the minimum at $\phi_1=0.5$ changes from cycle to cycle. A similar
depth of both minima suggests that temperatures of both components are
similar. The orbital period of this object is very long
($\approx38.2$~days) compared to other DPVs. The mean light curve is
typical for semi-detached or contact binary systems, thus, we suspect the
radius of the donor star is also one of the biggest among donors of known
DPVs or the total mass of the binary is smaller than for the other DPVs. We
suppose the changes in brightness of the components of the binary are in
the circumprimary disk or matter which left the gravitational potential of
the binary through $L_3$ point. The smaller amplitude of the longer cycle
during the eclipse near $\phi_1=0$ in Fig.~7 suggests that during that
eclipse the donor is closer to the observer and completely hides the
primary star and, at least, a substantial part of the disk. The donor does
not change its surface brightness and thus, close to $\phi_1=0$ we see
small dispersion of measurements around the fitted model. In the opposite
situation, when the primary component is closer to the observer, changes in
the brightness of the disk and probable matter close to $L_3$ point affect
the brightness measured by the observer. This changes cause the observed
dispersion of points seen in the light curve.

\subsection{OGLE-LMC-DPV-097 --  Deeper Primary Eclipses in Long Cycle \\
Minimum and Disappearing of Secondary Eclipses}
The light curve of this interesting DPV is shown in Fig.~8. The same
symbols in both panels correspond to the same epochs of observations. The
upper panel presents unfiltered data folded with the longer period and the
lower one -- light curve prewhitened with the longer cycle and folded with
\begin{figure}[htb]
\centerline{\includegraphics[width=9.5cm, bb= 20 25 565 745]{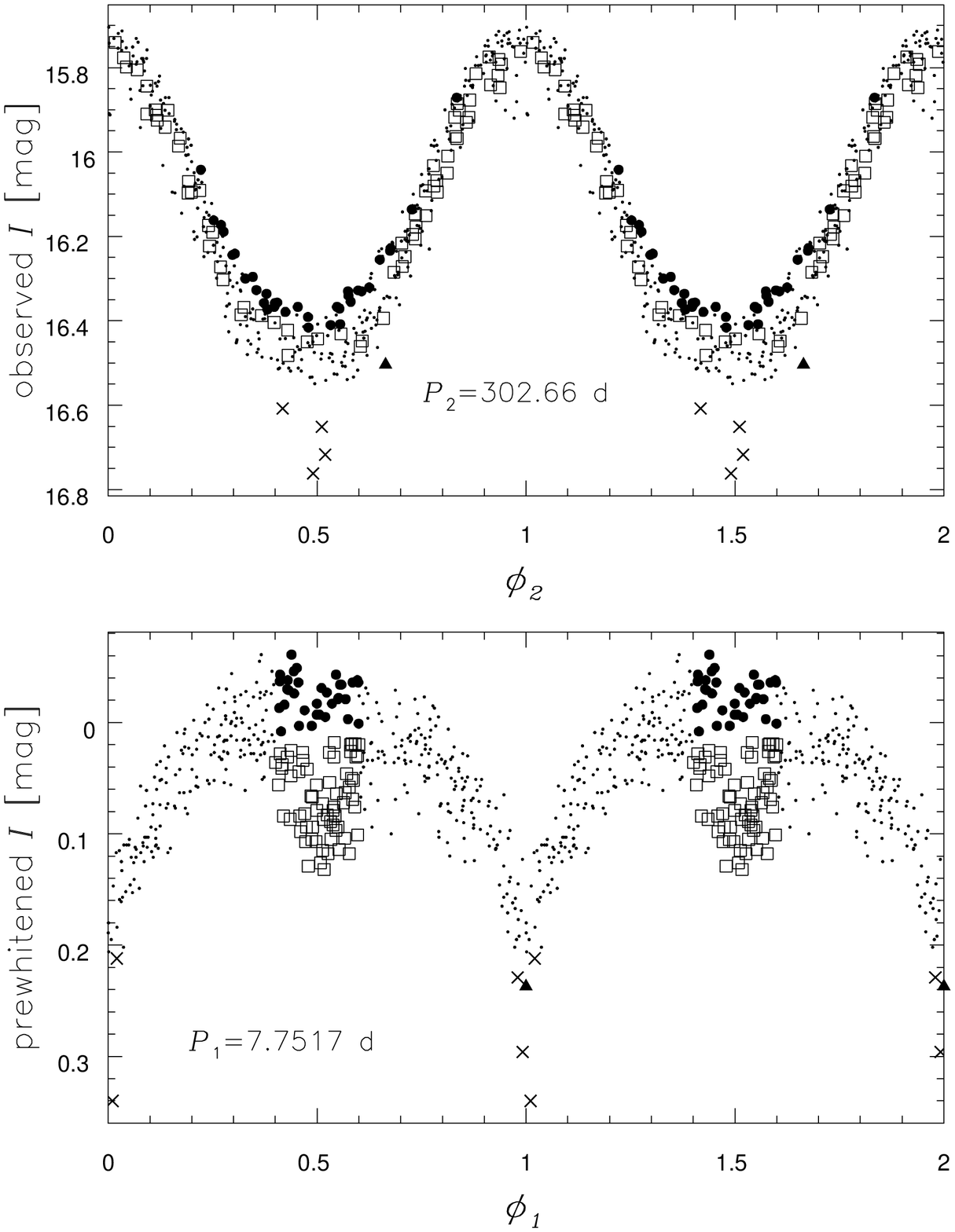}} 
\FigCap{Light curve of OGLE-LMC-DPV-097. {\it Upper panel} shows the 
unfiltered data phased with the longer period and {\it lower panel} shows
prewhitened data phased with the orbital period. For each epoch the same
symbols are used on both panels: empty squares are points from secondary
eclipse, filled circles correspond to measurements obtained at the same
phases of shorter period but when the secondary eclipse was not seen,
crosses mark four the faintest measurements of the longer cycle, filled
triangle -- third faintest point after prewhitening and all other
measurements are marked as dots.}
\end{figure}
the orbital period. On the upper panel one can see four points, which are
marked as crosses, clearly fainter than all the other. All of these points
are close to $\phi_2=0.5$. The same points are shown in the lower panel and
they turn out to be four out of five the faintest measurements and all of
them are in the middle of the primary eclipse. The third faintest point
(filled triangle) in the lower panel corresponds to the point at
coordinates $\phi_2=0.66$ and $I=16.50$~mag in the upper panel \ie it is
also close to the minimum light of the longer cycle and is fainter than
other measurements at the same phase of the longer cycle. During the
primary minimum the primary star and the disk around it are hidden behind
the secondary. The feature described above suggests that, at least for this
particular DPV, the long cycle originates near the circumprimary disk.

The other feature seen in Fig.~8 is a disappearing of the secondary minimum
of the orbital cycle. This phenomenon is seen only during the minimum light
of the longer cycle. To show this feature we divided the points in the
orbital phases between 0.4 and 0.6 into two groups using a horizontal line
not shown in Fig.~8. The points below this line (\ie during the secondary
minimum) are marked as empty squares and the ones above the line (\ie where
the secondary minimum is not observed) are marked with filled circles. When
we look at the distribution of these points on the upper panel we see that
almost all filled circles have brightness below mean brightness of the long
cycle. The empty squares are distributed almost uniformly during the long
cycle, however, it seems their number is smaller near the long cycle
minimum. The exact interpretation is hampered by the scatter of points
which remains after prewhitening with both cycles. We tried to differently
divide points between orbital phases 0.4 and 0.6 but no simple division
scheme neither gave better visualization nor showed additional features.

The naive explanation of disappearing of secondary eclipses may be the
change of the temperature ratio of both components of the binary. The
distribution of the filled circles suggests this ratio is closer to one
during the minimum than during the maximum of longer cycle. However, empty
squares observed during the minimum of the longer cycle contradict this
interpretation.

\subsection{OGLE-LMC-DPV-098 and OGLE-LMC-DPV-108 -- Deeper Primary \\
Eclipses in Long Cycle Minimum} We observe deeper primary eclipses during
the long cycle minimum not only for OGLE-LMC-DPV-097. Unfortunately, in
most cases the number of points fainter than the remaining measurements
with similar $\phi_2$ is small and thus inconclusive. The best examples of
stars showing deeper primary eclipses when $\phi_2\approx0.5$ are
OGLE-LMC-DPV-098 and OGLE-LMC-DPV-108, which light curves are shown in
Fig.~9 on the left and right panels, respectively. The top panels show
unfiltered light curves phased with $P_2$, middle ones -- light curves
prewhitened with the long cycle and folded with $P_1$ and bottom --
enlargement of the middle panel near $\phi_1=0$. The points outside the
primary eclipses (\ie $0.04<
\phi_1<0.96$) are marked with dots. For a better demonstration of the
effect we divided the points in the primary eclipses into two groups: ones
near $\phi_2=0$ (gray pentagons) and ones near $\phi_2=0.5$ (crosses). The
exact limits for crosses are $0.25<\phi_2<0.75$ for OGLE-LMC-DPV-098 and
$0.12<\phi_2<0.88$ for OGLE-LMC-DPV-108. This means that the separation
into crosses and gray pentagons is done using $\phi_2$ but lower panels of
Fig.~9 show that pentagons do not appear at the deepest locus of primary
eclipse.
\vspace*{-5pt}
\begin{figure}[htb] 
\centerline{\includegraphics[width=9.7cm, bb=20 35 565 745]{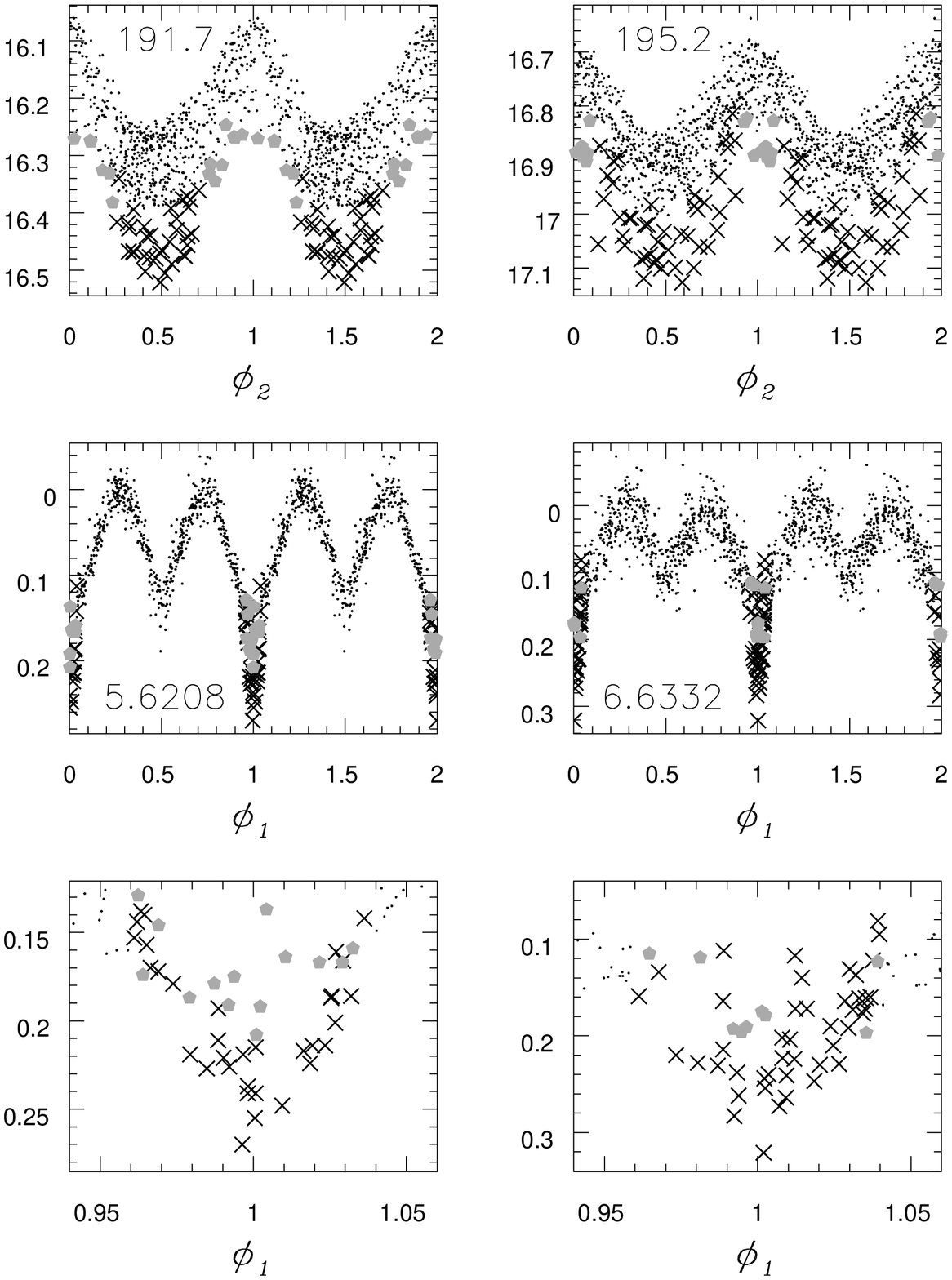}} 
\FigCap{Illustration of the larger amplitude of the primary eclipse 
during the minimum of the longer cycle for OGLE-LMC-DPV-098 ({\it left
panels}) and OGLE-LMC-DPV-108 ({\it right panels}). Values of $P_2$ and
$P_1$ are given in {\it upper} and {\it middle panels}, respectively. See
text for description.}
\end{figure}

Even though the criteria for the division of points in the primary eclipse
into two groups were selected manually, it is very likely that the effect
described above is not a random fluctuation of observations. We note that
the median uncertainties for points in the primary eclipse are 0.009~mag
and 0.013~mag for OGLE-LMC-DPV-098 and OGLE-LMC-DPV-108, respectively.

\subsection{OGLE-LMC-DPV-065 -- Period Changes}
Mennickent \etal (2005) called attention to the star LMC\_SC6\_57364 which
we designated as OGLE-LMC-DPV-065. They found out that between JD 2448800
and 2450000 the long term period was 340 days, while it shortened to 270
around JD 2450500. We tried to construct $O-C$ diagram for this star using
method described by Poleski (2008), however, the period changes of this
object are so rapid that it is hard to reliably construct the mean light
curve. In fact the light curve also changes its shape. We decided to divide
OGLE photometry of this object into several chunks and use
Schwarzenberg-Czerny (1996) method to measure periods separately in each
chunk. Gray vertical line in Fig.~1 presents the range of $P_2$ values
found by Mennickent \etal (2005) and in this study. We suspect the object
may move on the $P_1$--$P_2$ diagram much below the dashed line \ie where
no DPVs are observed. Values of $P_2$ as a function of time are presented
in Fig.~10 together with the linear fit to the data. The abscissa values
are average epochs in each of the chunks. The fit shown was used to
estimate average rate of period change which equals $\dd P_2/\dd t=
-0.01724\pm0.00040$. It is so rapid that if $\dd P_2/\dd t$ remains
constant the longer period will be 0 around year 2040. A few other DPVs
show obvious period changes with much smaller rate. Detailed investigation
of the period variability is beyond the scope of this paper.
\begin{figure}[htb]
\centerline{\includegraphics[height=9cm]{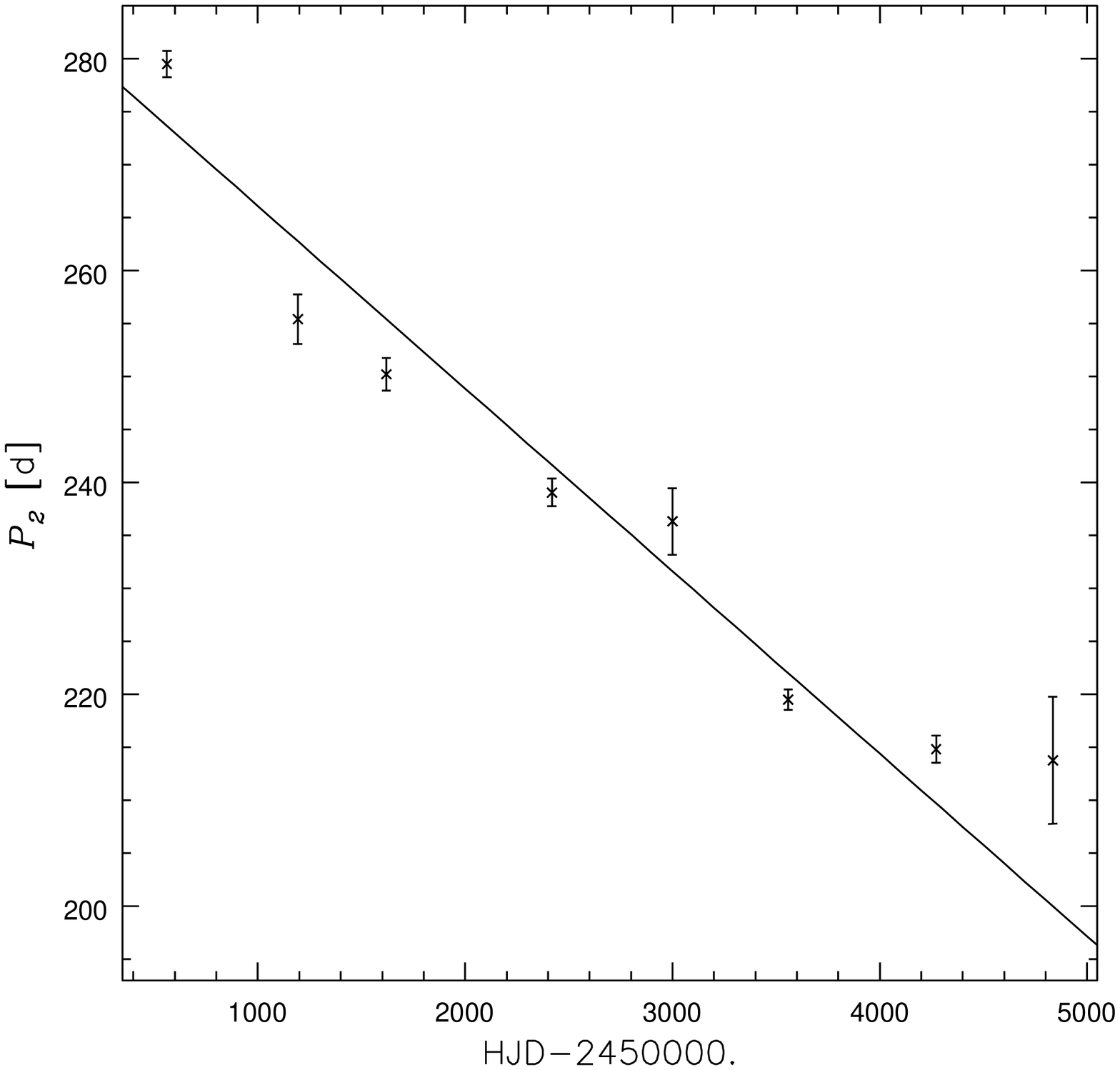}}
\FigCap{Period as a function of time for OGLE-LMC-DPV-065 during OGLE-II 
and OGLE-III observations.}
\end{figure}

\subsection{OGLE-LMC-DPV-054 -- Additional Changes in the Light Curve}
We also note that OGLE-LMC-DPV-054, similarly to OGLE-LMC-DPV-065, shows
rapid changes of the light curve morphology. The photometry prewhitened
with $P_1$ is shown in Fig.~11. The upper panel shows the {\it I}-band
magnitudes as a function of time and lower panels present data phased with
$P_2$ separately for OGLE-II (left) and OGLE-III (right). Although both
phased light curves show sinusoidal variability there are two episodes of
certainly non periodic behavior: $2450600<{\rm HJD}<2451350$ and since
${\rm HJD}=2454650$ up to the end of the data time-span. If the long cycle
variability is caused by circumbinary matter these two episodes may be
associated with variability of the rate of mass outflow. Presented light
curve does not cover the time after the second episode. Also the time-span
of observations before the first episode is smaller than $P_2$. We do not
see distinct changes of $\phi_2$ during the first episode.
\begin{figure}[htb]
\centerline{\includegraphics[height=11.5cm,angle=270]{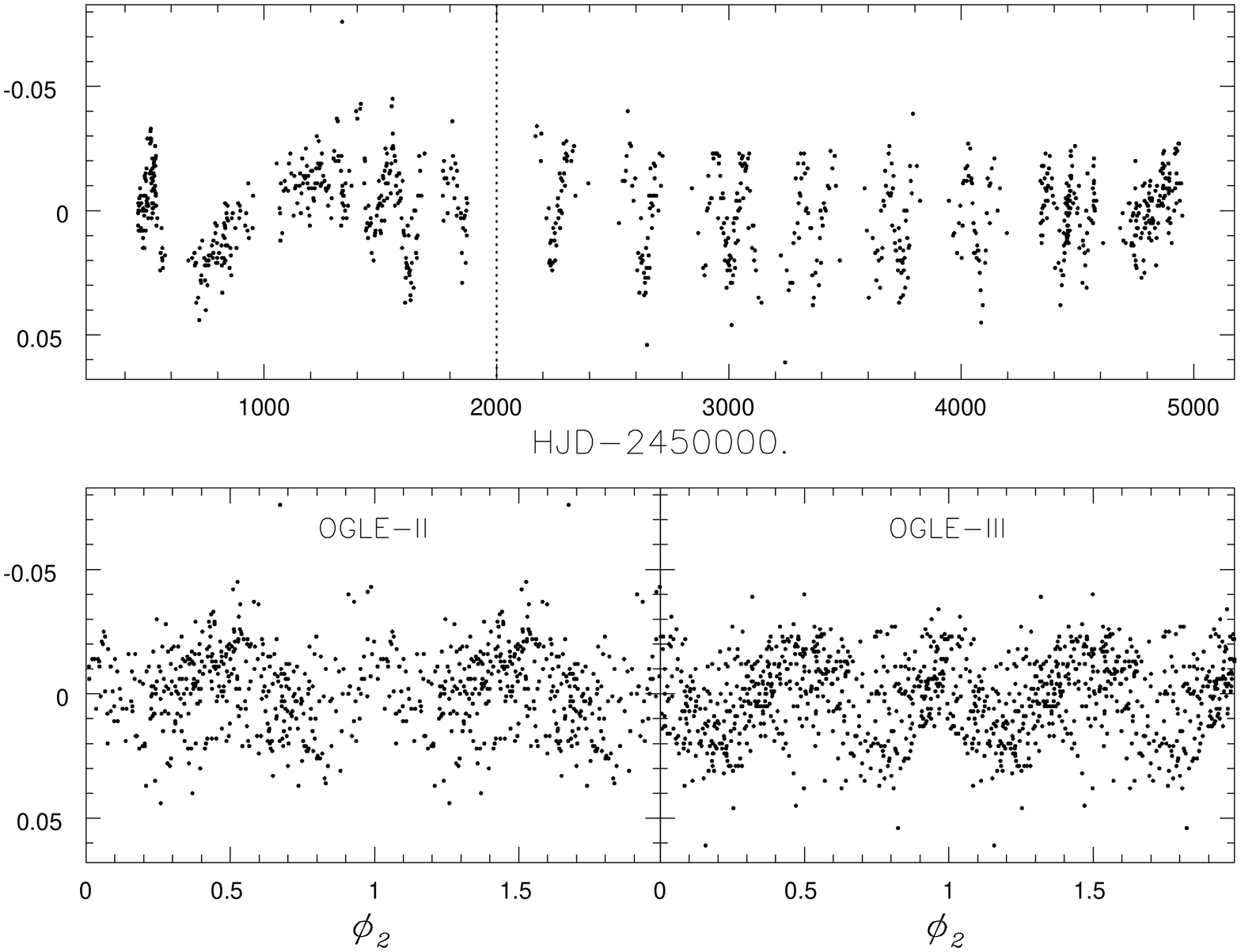}} 
\FigCap{Light curve of OGLE-LMC-DPV-054 prewhitened with 
$P_1=5.79598$~days. {\it Upper panel} shows photometric data without
folding. {\it Lower panels} show OGLE-II (${\rm HJD}<2452000$) and OGLE-III
(${\rm HJD}>2452000$) photometry folded with $P_2=254.7$~days.}
\end{figure}

\Section{Summary and Future Prospects}
We present the OGLE-III catalog of 125 DPVs in the LMC. It contains much
larger sample of these rare variable stars than known from all previous
studies. The extraordinary behavior of some of the presented DPVs needs
further observational efforts. Particularly interesting should be
OGLE-LMC-DPV-065 with its high period change rate. Currently conducted
fourth phase of the OGLE project is acquiring photometry of this object
which will be analyzed in future. We also plan to observe OGLE-LMC-DPV-097
with higher cadence during the minimum of the longer cycle in order to find
the shape of the secondary minimum and probably reveal the flux
contribution from different components. Color information can be obtained
from the archival photometry of MACHO and EROS projects each of which
simultaneously observed in two different bands. Color changes of
OGLE-LMC-DPV-054 during its erratic fluctuations may be very interesting.
Even without this additional observations our findings should be useful for
verification of hypothesis describing cause of long period variability.

\Acknow{
Authors are grateful Prof.~W.~Dziembowski for fruitful discussions and
R. Mennickent for reading the manuscript.  We thank Z.~Ko³aczko\-wski,
A.~Schwarzenberg-Czerny and J.~Skowron for providing the software used in
this study.

The research leading to these results has received funding from the
European Research Council under the European Community's Seventh Framework
Program\-me (FP7/2007-2013)/ERC grant agreement no. 246678. RP is supported
by the Foundation for Polish Science through the Start Program. This work
was also supported by the MNiSW grant NN203293533 to IS. The massive period
search was performed at the Interdisciplinary Centre for Mathematical and
Computational Modeling of Warsaw University (ICM UW), pro\-ject
no. G32-3. We are grateful to Dr.~M.~Cytowski for helping us in this
analysis.}


\begin{references}
\refitem{Alard, C.}{2000}{\AAS}{144}{363}
\refitem{Alard, C. and Lupton, R.H.}{1998}{\ApJ}{503}{325}
\refitem{Artyukhina, N.M., \etal}{1995}{~}{~}{``General Catalogue of Variable Stars'', 4rd ed., vol.V. Extragalactic Variable Stars, ''Kosmosinform'', Moscow} 
\refitem{Buchler, J.R., Wood, P.R., and Wilson, R.E.}{2009}{\ApJ}{703}{1565}
\refitem{Derekas, A., Kiss, L.L., and Bedding, T.R.}{2007}{\ApJ}{663}{249}
\refitem{Desmet, M., \etal}{2010}{\MNRAS}{401}{418}
\refitem{Djura\v{s}eviæ, G., Latkoviæ, O., Vince, I., and Ce\'eki, A.}{2010}{\MNRAS}{~}{in press}
\refitem{Mennickent, R.E., Pietrzyñski, G., D\'iaz, M., and Gieren, W.}{2003}{\AA}{399}{L47}
\refitem{Mennickent, R.E., Cidale, L., D\'iaz, M., Pietrzyñski, G., Gieren, W., and  Sabogal, B.}{2005}{MNRAS}{357}{1219}
\refitem{Mennickent, R.E., Ko³aczkowski, Z., Michalska, G., Pietrzyñski, G., Gallardo, R., Cidale, L., Granada, A., and Gieren, W.}{2008}{\MNRAS}{389}{1605}
\refitem{Mennickent, R.E., Ko³aczkowski, Z., Graczyk, D., and Ojeda, J.}{2010}{\MNRAS}{405}{1947}
\refitem{Pejcha, O., and Stanek, K.Z.}{2009}{\ApJ}{704}{1730}
\refitem{Pojmañski, G.}{1997}{\Acta}{47}{467}
\refitem{Poleski, R.}{2008}{\Acta}{58}{313}
\refitem{Schwarzenberg-Czerny, A.}{1996}{\ApJ}{460}{L107}
\refitem{Soszyñski, I., Poleski, R., Udalski, A., Szymañski, M.K., Kubiak, M., Pietrzyñski, G., Wyrzykowski, £., Szewczyk, O., and Ulaczyk, K.}{2008}{\Acta}{58}{163}
\refitem{Szymañski, M.K.}{2005}{\Acta}{55}{43}
\refitem{Udalski, A.}{2003}{\Acta}{53}{291}
\refitem{Udalski, A., Szymañski, M., Soszyñski, I., and Poleski, R.}{2008a}{\Acta}{58}{69}
\refitem{Udalski, A., Szymañski, M., Soszyñski, I., Kubiak, M., Pietrzyñski, G., Wyrzykowski, £., Szewczyk, O., Ulaczyk, K., and Poleski, R.}{2008b}{\Acta}{58}{89}
\refitem{Wo¼niak, P.R.}{2000}{\Acta}{50}{421}
\refitem{Wyrzykowski, £. \etal}{2009}{\MNRAS}{397}{1228}
\end{references}
\end{document}